\documentclass[reprint, amsmath, amssymb, aps]{revtex4-2}

\usepackage{amssymb}
\usepackage{graphicx}
\usepackage{dcolumn}
\usepackage{bm}
\usepackage{upgreek}
\usepackage{epstopdf}
\usepackage{extarrows}
\usepackage{amssymb}
\usepackage{amsmath}
\usepackage{color}
\begin{document}

\title{A Cyclical Route Linking Fundamental Mechanism and AI Algorithm: An Example from Tuning Poisson's Ratio in Amorphous Networks}

\author{Changliang Zhu$^{1,2,3}$, Chenchao Fang$^{1,2,4}$, Zhipeng Jin$^{1,2}$, Baowen Li$^{3}$, Xiangying Shen$^{3*}$, Lei Xu$^{1,2*}$}

\affiliation{$^1$ Department of Physics, The Chinese University of Hong Kong, Hong Kong, China\\
$^2$ Shenzhen Research Institute, The Chinese University of Hong Kong, Shenzhen 518057, China\\
$^3$ Department of Material Science and Engineering, Southern University of Science and Technology, Shenzhen, China\\
$^4$ Shenzhen Institute for Advanced Study, University of Electronic Science and Technology of China, Shenzhen 518110, China }

\email{Email: shenxy@sustech.edu.cn; xuleixu@cuhk.edu.hk}

\date{\today}

\begin{abstract}
{``AI for science'' is widely recognized as a future trend in the development of scientific research. Currently, although machine learning algorithms have played a crucial role in scientific research with numerous successful cases, relatively few instances exist where AI assists researchers in uncovering the underlying physical mechanisms behind a certain phenomenon and subsequently using that mechanism to improve machine learning algorithms' efficiency. This article uses the investigation into the relationship between extreme Poisson's ratio values and the structure of amorphous networks as a case study to illustrate how machine learning methods can assist in revealing underlying physical mechanisms. Upon recognizing that the Poisson's ratio relies on the low-frequency vibrational modes of dynamical matrix, we can then employ a convolutional neural network, trained on the dynamical matrix instead of traditional image recognition, to predict the Poisson's ratio of amorphous networks with a much higher efficiency. Through this example, we aim to showcase the role that artificial intelligence can play in revealing fundamental physical mechanisms, which subsequently improves the machine learning algorithms significantly.
}
\end{abstract}

\maketitle

\section{Introduction}
Using artificial intelligence (AI) to help scientific research has emerged as a prominent and well-recognized trend in the current academic community, shaping the future trajectory of scientific exploration \cite{wang2023scientific, stokes2020deep, tshitoyan2019unsupervised}. Fuelled by the significant advancements in computational science, machine learning has experienced unprecedented growth in recent years, showcasing various innovative algorithms and models, such as deep learning, reinforcement learning and active learning \cite{lecun2015deep, bellemare2020autonomous, tran2018active, jablonka2021bias}. These advancements are expected to significantly influence multiple dimensions of society, particularly within natural sciences, as they continue to evolve. Echoing Thomas Samuel Kuhn's observations in his famous book ``The Structure of Scientific Revolutions'', the iterative advancements in machine learning algorithms may catalyze a new wave of paradigm shifts in scientific research just as Copernicus proposed the heliocentric model.

AI has achieved notable progress in numerous scientific research domains \cite{karagiorgi2022machine, govorkova2022autoencoders, witman2023defect, dudte2023additive, chemrev.3c00708}. Such research traditionally begins with established theories and seeks tailored solutions for specific applications. It often involves data filtration and parameter optimization, where artificial intelligence demonstrates exceptional capabilities. Thus, the combination of machine learning and applied science appears very natural, fostering the rapid development of numerous achievements, such as using machine learning for protein structure prediction \cite{wan2020protein, yang2019machine, majewski2023machine}, employing various algorithms including deep neaural network \cite{pablo2023fast}, genetic algorithms \cite{chowdhury2020machine, wei2020genetic, song2022multi}, and Bayesian algorithms for metamaterial design \cite{hu2020machine, ju2017designing, ghahramani2015probabilistic}, enhancing density functional theory (DFT) through machine learning for predicting material properties \cite{fung2021machine,tsubaki2020quantum, liu2023supervised}, and constructing material genomes using big data and deep learning techniques \cite{ward2017atomistic, schmidt2019recent, kubevcka2023current}. These efforts expedite material screening and design, reducing the reliance on extensive experimental trial and error.

The conventional approach towards fundamental scientific research generally encompasses the following three stages:

1. Data Collection and Parameter Identification: This initial stage involves gathering pertinent data through observation and distinguishing crucial control parameters that influence the outcomes.

2. Mechanism Probing and Theoretical Modelling: The second stage delves into exploring the fundamental mechanisms at play. This is achieved through the application of both inductive and deductive reasoning to construct an encompassing theoretical model. This model, in turn, is designed to provide novel predictions.

3. Prediction Verification and Physical Control: The third stage involves the experimental verification of the newly formed predictions. Ultimately, the aim is to achieve an improvement on the understanding and control over the physical world based on the theory.

Within these three stages, machine learning algorithms typically can play a significant role in stage one but not in stages two and three. Consequently, the present role of AI in fundamental research primarily involves in supporting scientists in extracting data and control parameters from complex natural phenomena. For AI to be applicable in scientific research, the problems under investigation are typically very complex, challenging conventional research methods in making advancements. This complexity usually arises from numerous data or parameters, which impede the capture of key factors for constructing a theoretical model and thus has to rely on machine learning's computational power. However, analogous to stage 3 above, machine learning analysis may also gain significant improvement from the fundamental understanding obtained in stage 2, which has rarely been demonstrated before. In this study, we demonstrate such a three-stage ``AI for science'' cyclical route: for the Poisson's ratio in disordered networks, we first extract crucial tuning approaches with machine learning, then probe the fundamental mechanism based on the machine learning discoveries, and at last improve the machine learning algorithms based on the underlying mechanism. Our study thus provides an excellent example of three-stage machine learning analysis, reveals the fundamental mechanism of Poisson's ratio with machine learning, and gives a novel design framework to achieve arbitrary Poisson's ratio from machine learning.

Previous studies have revealed that in amorphous elastic networks composed by bonds and nodes, extreme Poisson's ratios can be achieved by removing specific bonds, potentially leading to auxetic behavior \cite{lakes1987foam, hexner2018role, liu2019realizing}. However, the relationship between the Poisson's ratio of such a complex system and its structure remains a subject of ongoing debate. This ambiguity stems from the non-affine nature of amorphous networks, where, under low connectivity or nonuniform nodes locations, the displacement field of internal nodes tends to chaotic under external stress or load \cite{zaccone2014short, schlegel2016local}. Analyzing such a chaotic displacement field is challenging due to network complexity. Moreover, the presence of non-affinity leads to a redistribution of the bond contribution to the system's Poisson's ratio as bonds are removed. Consequently, there is no optimal bond-cutting strategy to adjust system Poisson's ratio across the theoretical limits.

In the investigation of the aforementioned research problem, to obtain a more general understanding of the correlation between the Poisson's ratio of amorphous networks and their structural configuration, the initial step is to obtain the optimal bond pruning strategy corresponding to a specific Poisson's ratio value, i.e., removing the minimum number of connections to achieve the desired Poisson's ratio value. Hence, we employed a machine learning technique known as the simulated annealing algorithm to tune the Poisson's ratio across theoretical  limits (-1 and 1). Unexpectedly, networks displaying an extremely negative Poisson's ratio and those with an extremely positive Poisson's ratio exhibit almost indiscernible structural features to the naked eye, indicating that the machine learning algorithm discovers a new class of auxetic materials. Subsequent theoretical analysis reveals a direct connection between the Poisson's ratio and network's lowest-frequency normal modes from dynamical matrix. Building upon this theoretical insight, we trained a Convolutional Neural Network (CNN) based on the dynamical matrix to predict Poisson's ratio. Unlike traditional CNN models intended for image recognition and handling vast amounts of pixel data, the data volume associated with the dynamical matrix is significantly smaller. Consequently, this approach drastically reduces the computational power required for Poisson's ratio prediction.

\begin{figure*}[htpb]
\begin{center}
\centerline{\includegraphics[width=\linewidth]{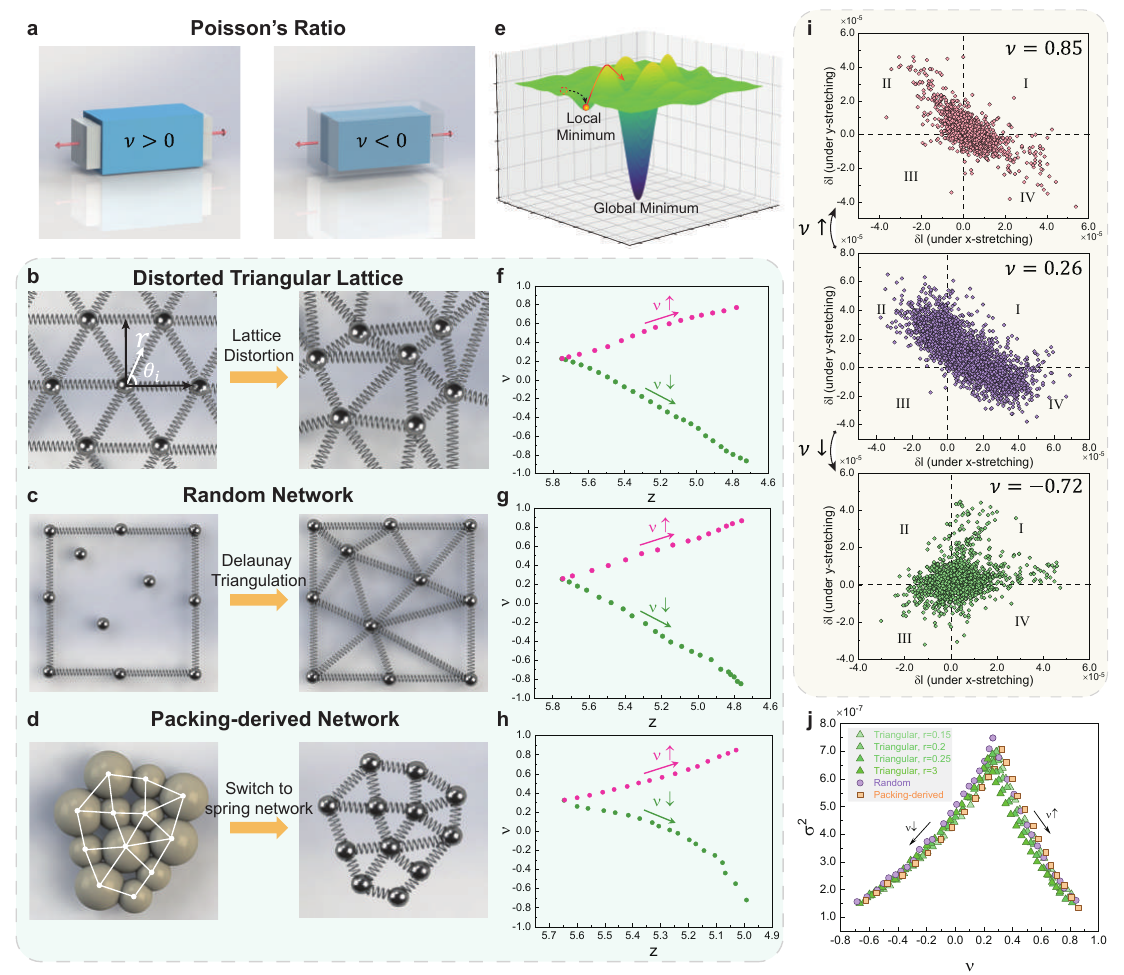}}
\caption{\label{fig:1}
Tuning $\nu$ from positive to negative in three types of amorphous networks. \textbf{a}, a schematics showing the positive and negative $\nu$ behaviors. The red arrows indicate the uniaxial stretching, the original state is drawn in blue, and the deformed state is drawn in light gray. For $\nu>0$, the longitudinal direction elongates while the transverse direction shrinks; for $\nu<0$, both directions elongates. \textbf{b-d}, the construction of three types of amorphous networks: the distorted triangular lattice, the Delaunay-triangulated random network, and the packing-derived network. \textbf{e}, a schematics demonstrating local minimum and global minimum. \textbf{f-h}, tuning towards either positive or negative Poisson's ratio from the same initial state, for the three types of amorphous networks shown on the left. Note that all configurations throughout tuning contain only convex polygons. \textbf{i}, the statistics of single-bond length changes under stretching. Each data point represents one single bond: its x and y coordinates are the length changes under x and y stretching respectively. The middle panel is the original system, and the top and bottom panels are the systems after $\nu$-increase and $\nu$-decrease tunings respectively. Clearly, systems with positive $\nu$ and negative $\nu$ exhibit distinct data distributions: the former are mostly in quadrants II and IV while the latter are mostly in quadrants I and III. \textbf{j}, the variance of data points similar to \textbf{i} is plotted against $\nu$, and a reasonable collapse is observed.}

\end{center}
\end{figure*}

\section{Results}

\subsection{Tuning Poisson's ratio with simulated annealing algorithm}

In our amorphous system, we utilize a network model composed of $N$ nodes connected by $N_b$ bonds, and each bond is an ideal massless spring with the spring constant $k$ \cite{mao2018maxwell,lubensky2015phonons, zhang2009thermal, souslov2009elasticity, lera2018topological, nie2017role}. The coordination number $z=2N_b/N$ represents the average number of bonds per node \cite{rocklin2021elasticity,zhou2019topological,rocklin2017transformable,rocklin2016mechanical,wyart2008elasticity, zaccone2011approximate}. By removing a small fraction of bonds in the network, one can tune the system's Poisson's ratio from positive to negative\citep{goodrich2015principle, shen2021achieving,PhysRevApplied.18.054052} as shown in Figure 1a. To ensure the general validity of this finding, we construct three distinct types of amorphous networks and verify the universality of our results in all of them. These network types include: (1) the distorted triangular lattice (Figure 1b), (2) the Delaunay-triangulated random network (Figure 1c), and (3) the network derived from packing configurations (Figure 1d).

We use the simulated annealing algorithm (see Methods for details), an optimization method (see Figure 1e) that employs an oscillation-like mechanism to escape from local optima solutions, to approach the global optimum. After calculation, we have achieved a tuning range of $-0.9<\nu<0.9$, which approaches the theoretical limit of $-1<\nu<1$ in 2D isotropic systems.To ensure orientation independence, we separately determine $\nu_x$ and $\nu_y$ under x-stretching and y-stretching, respectively. The average of these values, $\nu=(\nu_x+\nu_y)/2$, represents the system's overall $\nu$. Note that due to the isotropic nature of our system, $\nu_x$ and $\nu_y$ exhibit similar values and behaviors. As depicted in Figure 1f, g, h, our machine learning algorithm successfully achieves $-0.9<\nu<0.9$ in all three types of amorphous networks, thereby demonstrating its remarkable tuning range and applicability.

To visualize the changes occurring at the single-bond level during the tuning process, we analyze the length changes of each individual bond under $x$-stretching and $y$-stretching. The resulting statistics are presented in Figure 1i: the middle panel depicts the initial configuration, while the upper and lower panels showcase the situations after $\nu$-increase and $\nu$-decrease tunings, respectively. In these plots, each point represents a bond, with its $x$-coordinate and $y$-coordinate indicating the length changes under $x$-stretching and $y$-stretching, respectively. It is evident that the original data points in the middle plot are broadly dispersed across quadrants II and IV, a result of opposite deformation signs due to the positive Poisson's ratio. However, after either increasing or decreasing $\nu$, the data points converge towards the center, signifying that both types of tuning reduce the extent of bond length changes during stretching, with only a minor fraction undergoing substantial transformations. Intriguingly, tuning $\nu$ to a negative value causes a redistribution of the points into quadrants I and III, in stark contrast to the positive-$\nu$ situations portrayed earlier. Given that quadrants I and III contain $x$ and $y$ deformations with the same sign, this behavior aligns well with the auxetic phenomenon, where both dimensions simultaneously experience expansion or shrinkage.

Hence, Figure 1i clearly illustrates that distinct distribution patterns of bond variations under load correspond to different $\nu$ values. To quantify these patterns, we calculate the variance of each distribution, denoted as $\sigma^2={\sum_{i=1}^{N}(\vec{r_i}-{\vec{r}_{mean}})^2}/{N}$, for both of the $\nu$-decrease and $\nu$-increase tunings. Interestingly, when these $\sigma^2$ values are plotted against $\nu$, data from various amorphous systems exhibit reasonable collapse, as depicted in Figure 1j. This manifestation indicates a close correlation between the statistical variance of single-bond variations under load, $\sigma^2$, and $\nu$. Moreover, our machine learning approach effectively re-distributes $\sigma^2$ into quadrants I and III, thereby achieving negative-$\nu$ systems.

\begin{figure*}[htpb]
\begin{center}
\centerline{\includegraphics[width=\linewidth]{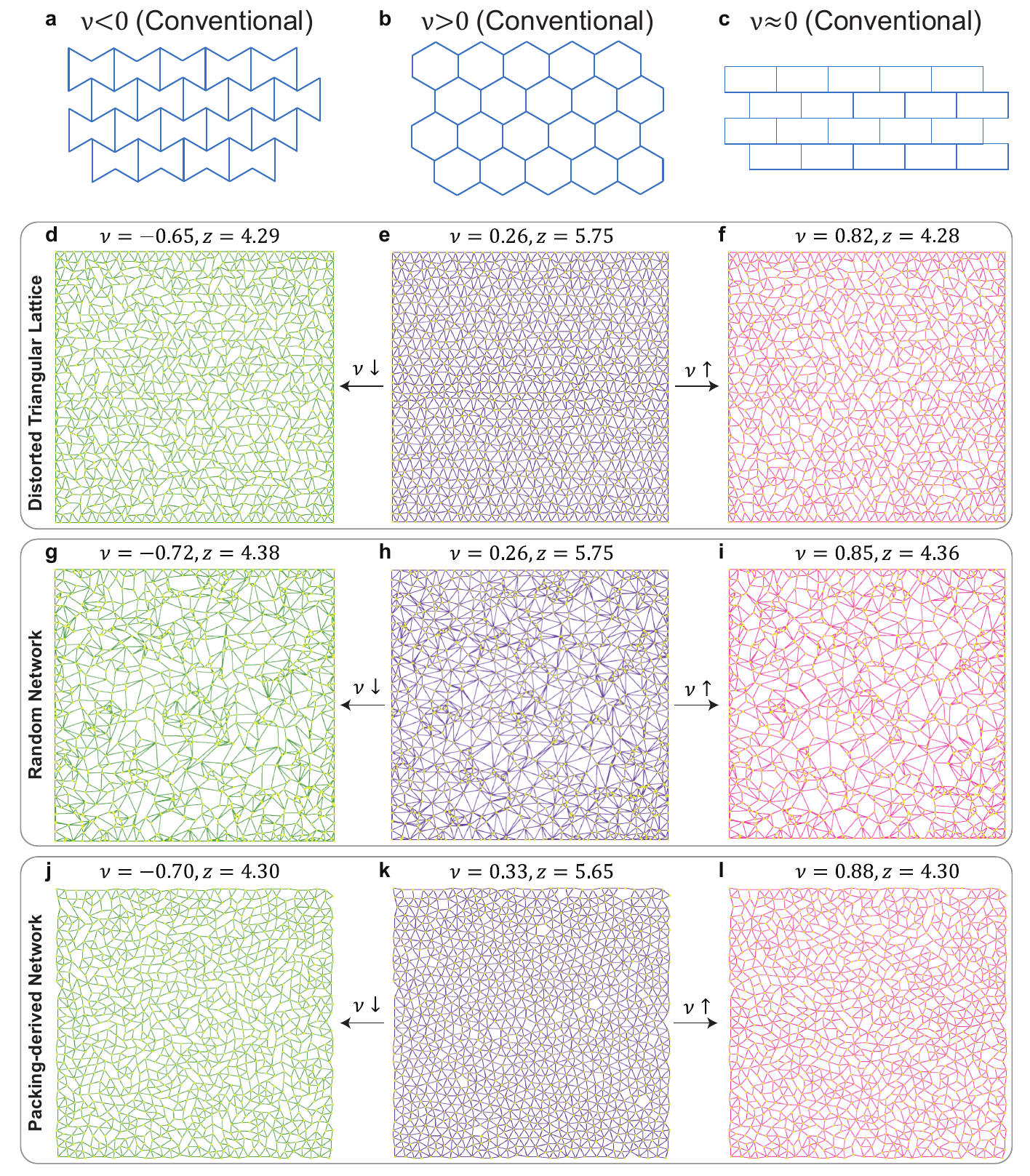}}
\caption{\label{fig:2}
The classical auxetic structures are fundamentally different from the auxetic structures discovered by our machine learning. \textbf{a-c},the three classical structures, characterized by $\nu<0$, $\nu>0$, and $\nu \approx 0$, are derived from the fundamental bow-tie structure. \textbf{d-l}, the three rows correspond to the three different amorphous systems. Within each row, the middle panel shows the original structure, and the left and right panels show the negative and more positive configurations found by the machine learning. Apparently, in the left panels of auxetic structures there is no concave structures typically existing in the classical auxetic structures. Also note that the negative-$\nu$ panels on the left are structurally very similar to the positive-$\nu$ panels on the right with almost a same $z$.
}
\end{center}
\end{figure*}

Next, we elucidate the fundamental difference between our machine learning generated structures and conventional designs in Figure 2. In the first row, we show the three conventional designs with Poisson's ratio $\nu<0$, $\nu>0$ and $\nu \approx 0$, all based on the typical bow-tie structures ~\cite{evans1991molecular,bronder2022design, buckmann2012tailored}. Rows 2 to 4 demonstrate our amorphous networks generated from a distorted triangular lattice (row 2), Delaunay-triangulated random network (row 3), and packing-derived network (row 4). In the middle of each row, we present the original networks with $\nu$ around 0.3, while the highly-negative and highly-positive networks resulting from bond cutting are displayed on the left and right panels, respectively. Notably, our auxetic structures in the left panels differ significantly from the conventional designs in Figure 2 \textbf{a}: (1) our structures are amorphous and isotropic in all directions, whereas the conventional designs exhibit directional-dependent responses; (2) our structures consist entirely of convex polygons (see the bond-angle distribution in SI for direct proof), while the conventional structures depend on concave structures to realize negative Poisson's ratio. Thus it is possible to realize auxetic behavior with all-convex structures and concave structures are not always required. Additionally, we have conducted computations on varying system sizes, ranging from $12 \times 12$ nodes to $32 \times 32$ nodes. The results consistently demonstrate the tunability of Poisson's ratio (see Supplementary Figure 2 and Figure 3). It is evident that regardless of the system size, the Poisson's ratio converges consistently towards -0.9 for auxetic networks and towards 0.9 for more positive networks. This observation underscores the consistent tunability of our machine learning algorithm across different system dimensions and scales.

Interestingly, the structures with highly negative and highly positive $\nu$ appear quite similar. Both exhibit isotropic structures with similar connectivity (i.e., similar $z$) and bond angle distribution (see SI), as depicted in the left panels versus the right panels in Figure 2. However, when subjected to external loading, these visually similar structures exhibit completely opposite Poisson's ratios. This high structural similarity between structures with positive and negative $\nu$ has not been previously reported, further demonstrating the uniqueness and novelty of the machine learning-generated structures. By combining these visually similar structures, one can also create functionally gradient materials without structural gradient, as illustrated in the Supplementary Information.

\subsection{Experimental confirmation of designed amorphous networks}

We have successfully achieved continuous tuning of $\nu$ from highly positive to highly negative by employing the machine learning optimization algorithm in general amorphous networks. To validate the simulated annealing algorithm is practical reliable, we experimentally build the networks designed by machine learning and compare their Poisson's ratio against the numerical prediction. We 3D print amorphous elastomer networks with negative, positive, and zero $\nu$. Note that due to experimental convenience, the 3D printed structures have a smaller system size compared to the simulation structures in Figure 2. However the machine learning protocol used for both the experiment and simulation is exactly the same.

In our experiments, we apply a compression strain of $\varepsilon = -0.1$ to each network. We then measure the corresponding deformation field and Poisson's ratio using optical imaging. For practical purposes, the experiment is designed to apply only compressive strain to prevent the risk of bond breakage and potential experimental failure that could arise from the application of extensional strain. However, we have numerically tested extensional strain and obtained consistent results (see Supplementary for details). To validate the robustness of our design, we test all three types of amorphous networks, which are depicted in the three rows of Figure 3b-j. Remarkably, for each type, we successfully achieve negative, positive, and nearly-zero $\nu$ values. The left panels demonstrate shrinkage or negative $\nu$ in the horizontal response, the right panels show expansion or positive $\nu$, and the middle panels exhibit almost no change or zero $\nu$. The red boxes indicate the initial positions of the systems. The exact deformation process can be observed in Movie-1 to Movie-9. Note that the achieved tuning range is $-0.6 < \nu < 0.5$, which is narrower than the simulation range of $-0.9 < \nu < 0.9$. However, this range still confirms that simulated algorithm is quite useful in tuning the Poisson's ratio. One key reason for this discrepancy is that the simulation assumes a pure spring interaction, whereas the bending effect inevitably exists in actual bonds \cite{feng2016nonlinear, reid2018auxetic, rocks2017designing, broedersz2011criticality, reid2019ideal, rocklin2014self}. Another reason is due to the static friction at the sample's top and bottom boundaries, which limits the horizontal motion and leads to a reduced tuning range (see Supplementary for more details from simulation). Despite this inherent bending complexity, all amorphous systems demonstrate a significant tuning range, thus highlighting the robustness and general validity of our machine learning design.

\begin{figure*}[htpb]
\begin{center}
\centerline{\includegraphics[width=\linewidth]{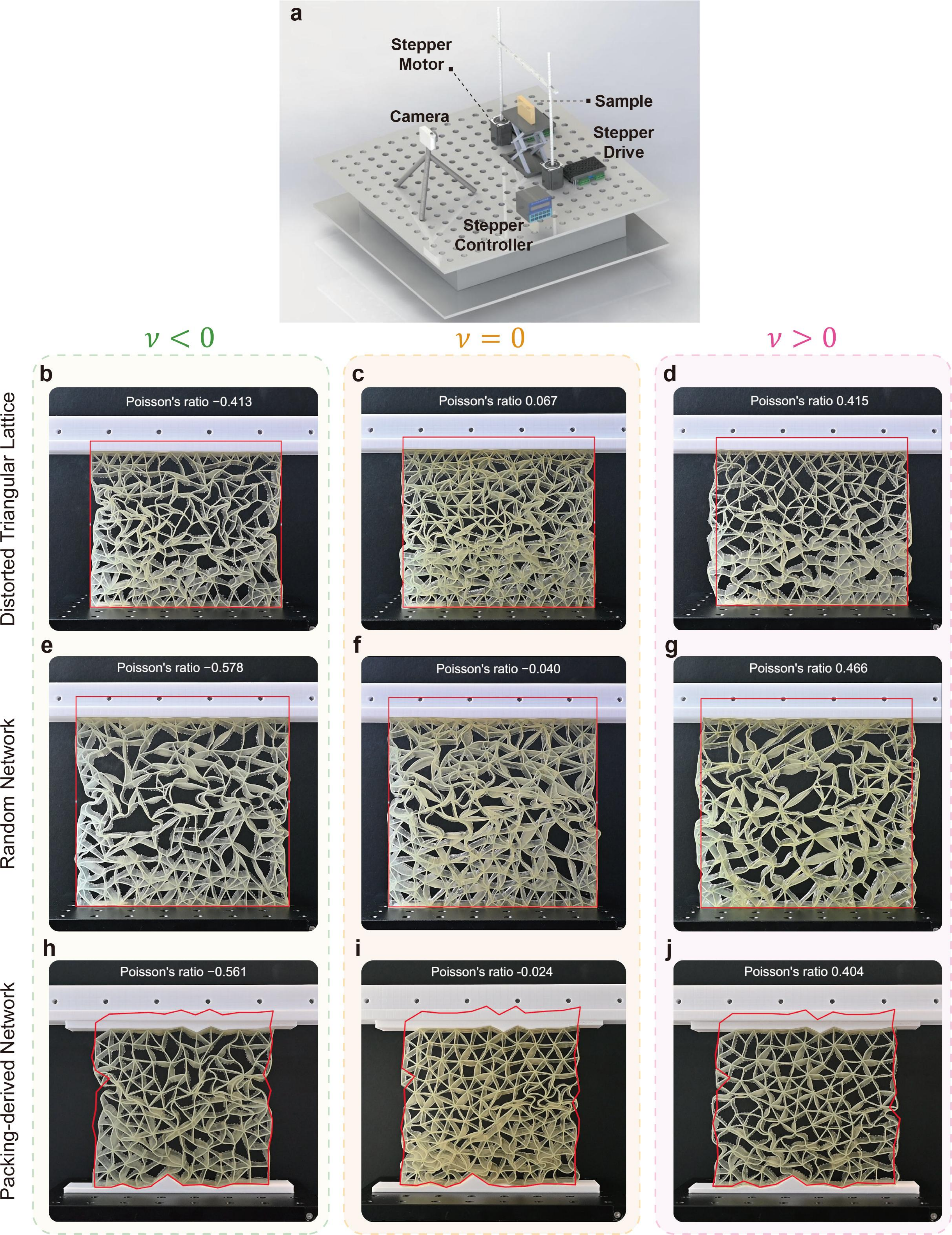}}
\caption{\label{fig:3}
Experimental realization with 3D-printing. \textbf{a}, experimental setup for $\nu$ measurement. The sample is under a compression loading on its top surface driven by two step motors. Compression strain and speed are precisely controlled by the motors. The Poisson’s ratio is measured from the area change of sample during the compression, which is recorded by a camera in front of the sample. \textbf{b}-\textbf{j}, the three rows show the situations of the three amorphous systems. Each snapshot is taken at the compression strain of  $\varepsilon = -0.1$, and the red boxes indicate the original boundary without loading. Clearly, the left panels exhibit a negative $\nu$, the middle panels exhibit a nearly-zero $\nu$, and the right panels exhibit a positive $\nu$. (Multimedia view).
}
\end{center}
\end{figure*}

\subsection{Poisson's ratio determined by a few normal modes}

The fundamental origin that determines $\nu$ remains an open question, which we now elucidate with the normal mode picture. In all initial networks with high-z values (close to 6 in 2D) and the subsequent ones generated from our optimization algorithm, we find that only a few low-frequency vibrational normal modes (typically one to two) can determine their Poisson's ratio. For a 2D amorphous network with $N$ nodes, we construct a $2N\times 2N$ dynamic matrix and calculate its $2N$ eigenvalues and eigenmodes, corresponding to the vibrational frequencies and normal modes \cite{mao2010soft, bertoldi2017flexible, zou2009packing, ridout2022correlation, mizuno2016elastic}. Across the broad range of $-0.9<\nu<0.9$ in all our generated systems, we observe that $\nu$ is universally determined by one to two vibrational normal modes.

To characterize the Poisson's ratio with normal modes, we define an effective Poisson's ratio, $\nu'$, for each mode. As depicted in Figure 4a and Figure 4b, when we treat the vibrational vectors of a normal mode as displacement vectors on the nodes, the original structure becomes `deformed', resulting in an effective Poisson's ratio, $\nu'$, for that mode. A mode may exhibit either a positive or a negative $\nu'$ (see Figure 4a,b). We demonstrate that the actual $\nu$ is determined by the superposition of a few important modes' $\nu'$.

To identify these important modes, we project the actual deformation under load onto all the normal modes, and analyze the projection probabilities or weights. Since normal modes are orthogonal to each other and form an orthonormal set of basis consisting of all $2N$ modes, we can project any actual deformation field onto this set of basis:

\begin{equation}
|\delta r\rangle = \sum_{i = 1}^{2N} C_{\omega i}|\omega_i\rangle
\end{equation}

\noindent Here $|\delta r\rangle$ is the actual deformation field under an external load, which is a $2N\times1$ dimensional vector normalized to unit amplitude. $|\omega_i\rangle$ is the $ith$ normal mode with the frequency $\omega_i$. $C_{\omega i}=\langle\omega_i|\delta r\rangle$ is the projection pre-factor of $|\delta r\rangle$ on mode $|\omega_i\rangle$, and $|C_{\omega i}|^2$ has the physical meaning of projection probability: it gives the weight or importance of mode $|\omega_i\rangle$ in the actual deformation field $|\delta r\rangle$. To obtain a result independent of direction, the external load is applied separately in x and y directions, and the average is obtained as the final projection probability.

Therefore, the importance or weight of each mode in the actual deformation can be illustrated by plotting $|C_{\omega i}|^2$ against $\omega_i$ in Figure 4c. Notably, two prominent peaks can be observed in the low-frequency range, revealing the significance of two modes, while other modes play a negligible role (note that the weight is zero at high frequencies, and only the low-frequency range is meaningful). Interestingly, these two modes typically exhibit opposite effective Poisson's ratios $\nu'$: one demonstrates a highly positive $\nu'$ while the other exhibits a highly negative $\nu'$ (denoted as $\omega^+$ and $\omega^-$ respectively in Figure 3c). By considering their superposition based on their weights, the actual Poisson's ratio $\nu$ can be determined as $\overline{\nu'}=|C_{\omega+}|^2 \nu'(\omega^+) + |C_{\omega-}|^2 \nu'(\omega^-)$. This relationship is validated by the linear relation depicted in Figure 4d.

As a result, in our systems the origin of $\nu$ essentially originates from the $\nu'$ of normal modes. Certain low-frequency modes possess either positive or negative $\nu'$. The external load significantly excites one or two such modes. The superposition of these $\nu'$ values governs the actual $\nu$ of the system (see Supplementary for a detailed derivation). This mechanism applies to all three types of amorphous systems we generated, including the distorted triangular lattice, random network, and packing-derived network, as demonstrated by the excellent data collapse depicted in Figure 4d.

The fundamental origin of $\nu$ from the $\nu'$ of normal modes provides a clearer understanding of the tuning process. Figure 4e illustrates how the competition between the two important modes, $\omega^+$ and $\omega^-$, determines the actual value of $\nu$. When we tune $\nu$ from its original value of 0.3 to a negative value, the importance of $\omega^-$ increases, eventually becoming dominant, while the significance of $\omega^+$ decreases, becoming negligible. This results in the effective reduction of $\nu$ from a positive to a negative value. Additionally, $\omega^+$ and $\omega^-$ initially approach each other, then overlap, and finally separate, as depicted in the top to bottom panels. Notably, when these two modes overlap, they exhibit the same height, leading to the cancellation of positive and negative $\nu'$. Consequently, a nearly-zero value of $\nu$ is obtained (see panel 3). In contrast, when we tune $\nu$ from its original value of 0.3 to a more positive value, the opposite trend is observed in Figure 4f. The two peaks continue to separate, and the height of $\omega^+$ increases while the height of $\omega^-$ decreases, eventually becoming negligible. The increasing weight of $\omega^+$ contributes to the overall increase in the actual value of $\nu$. Also note that $\omega^-$ and $\omega^+$ exhibit a linear dependence with respect to $z-z_c$, with $z_c=4$ being the isostatic point in 2D (see SI for the data plot).

\begin{figure*}[htpb]
\begin{center}
\centerline{\includegraphics[width=0.9\linewidth]{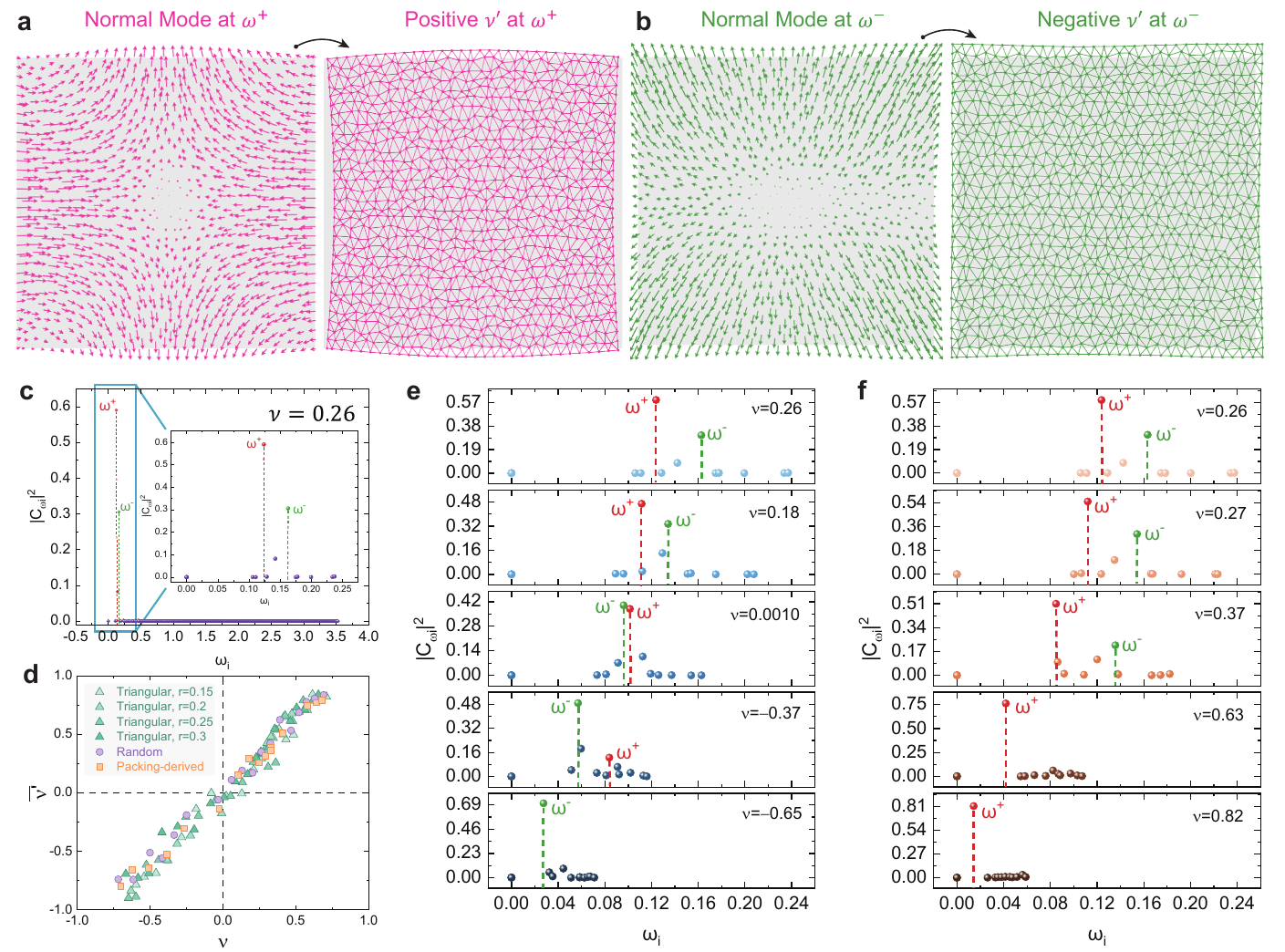}}
\caption{\label{fig:4}
The effective Poisson's ratio of normal modes, $\nu'$, determines the actual $\nu$ of the system. \textbf{a}, the left panel shows the normal mode at the frequency $\omega^+$, and the gray area indicates the original state. The right panel shows the configuration after adding the vibrational vectors: clearly the network expands in y direction and shrinks in x direction, which is a typical feature of positive Poisson's ratio. \textbf{b}, The normal mode at the frequency $\omega^-$ shows simultaneous expansion in both x and y directions, which is a typical auxetic behavior. \textbf{c}, $|C_{\omega i}|^2$ shows the weight or importance of different normal modes in an actual stretching deformation. Only the low-frequency range is important and the values at high frequencies are all zero. Clearly there are two peaks in the inset: one at $\omega^+$ with positive $\nu'$ and one at $\omega^-$ with negative $\nu'$. The superposition of the two $\nu'$ based on their weights, $\overline{\nu'}=|C_{\omega+}|^2 \nu'(\omega^+) + |C_{\omega-}|^2 \nu'(\omega^-)$, determines the actual $\nu$. \textbf{d}, $\overline{\nu'}$ from normal modes versus actual $\nu$ shows a nice linear relation, and an excellent data collapse across various systems is observed. Clearly, one to two normal modes can universally determine the actual Poisson's ratio across various amorphous systems. \textbf{e}, from top to bottom, as $\nu$ is tuned to negative values, the weight of $\omega^-$ keeps increasing to dominant while the weight of $\omega^+$ keeps decreasing to negligible. When the two modes have similar weights, their $\nu'$ cancels out and the system exhibits a nearly-zero $\nu$ (see the middle panel). The positions of two peaks first approach each other, then overlap and eventually separate apart. \textbf{f}, from top to bottom, as $\nu$ is tuned to more positive values, the weight of $\omega^+$ keeps increasing to dominant while the weight of $\omega^-$ keeps decreasing to negligible. The two peak positions keep separating apart.
}

\end{center}
\end{figure*}

\begin{figure*}[htpb]
\begin{center}
\centerline{\includegraphics[width=0.9\linewidth]{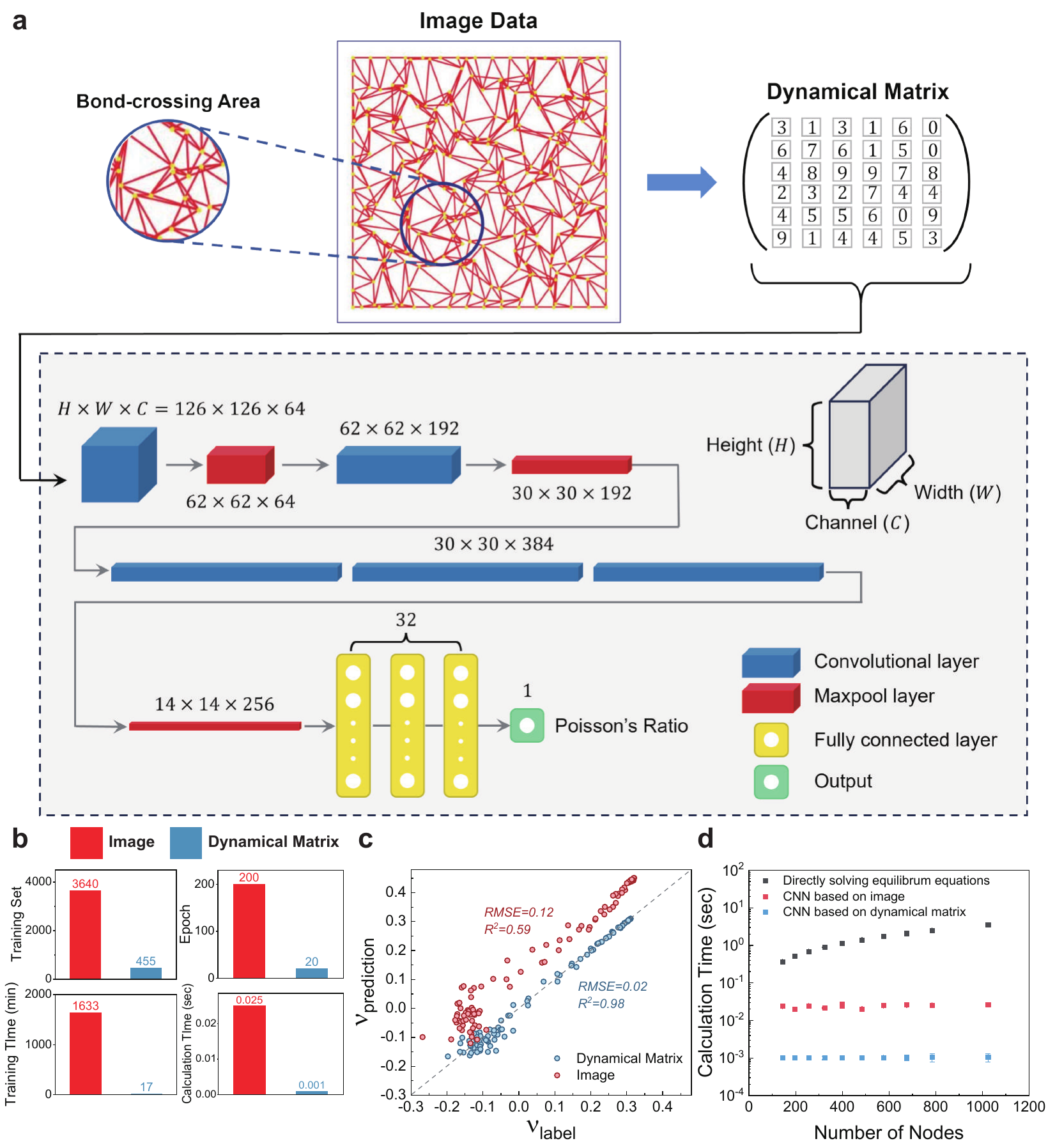}}
\caption{\label{fig:5}
Scientific discovery serves as a catalyst for advancements in deep learning. \textbf{a}, The presence of bond-crossing areas poses challenges for deep learning models, as neural networks struggle to discern whether two crossed bonds share a common node. This challenge is mitigated by substituting images with the dynamical matrix as input, as illustrated by the architecture of AlexNet. \textbf{b}, Comparisons of the training results of CNN models using images and dynamical matrices as inputs. \textbf{c}, Comparing the prediction performance of Poisson's ratio from training based on dynamical matrix and images as the testing dataset, which demonstrates a much better performance from training based on dynamic matrix. \textbf{d}, Calculation of time cost by directly solving equilibrium equations and CNN models based on images and dynamical matrices. The comparison is conducted across amorphous networks of different sizes, ranging from $12 \times 12$ to $32 \times 32$.
}
\end{center}
\end{figure*}

\subsection{Fundamental mechanism improves deep learning algorithm}

The underlying mechanism governing the Poisson's ratio, as revealed by our normal mode analysis, can be effectively utilized to enhance the efficiency and broaden the prediction range of deep learning models. Traditional CNN models rely on image-based training data, which often demand extensive datasets and result in prolonged training time. Moreover, these models may face challenges in situations where image datasets are unsuitable. This is evident in scenarios such as distorting nodes of a triangular lattice to induce auxetic behavior, where multiple bond-crossing areas may emerge, as depicted in Figure 5a. This poses a challenge for CNN models, as they may mistakenly put a node at the crossing point of two bonds whereas there is no node in reality. Analogous to the traditional image input that is a matrix of pixels, we recognize that the mechanical behavior is encoded within network's dynamical matrix and thus propose to use its dynamical matrix as input to train deep learning models. To demonstrate this concept, we trained the AlexNet model using the dynamical matrix as the training dataset.

The training and prediction performance is further compared with a ResNet-50 model trained using image datasets, as illustrated in Figure 5b-d. Figure 5b highlights that a well-trained network utilizing the dynamical matrix accelerates prediction time by 25 times compared to models using images as inputs. Training with images is notably more challenging, requiring a dataset nine times larger and ten times more training epochs, resulting in a training time 96 times longer. As depicted in Figure 5c, the AlexNet demonstrates high accuracy in predicting Poisson's ratio with the input dynamical matrix, achieving an $RMSE$ (root mean square error) value of $0.02$. The scatter plots closely align around the grey dashed line representing the curve of $\nu_{prediction} = \nu_{label}$. However, the prediction accuracy of ResNet-50 trained on images are much worse, with an $RMSE = 0.12$, due to the difficulty of telling whether or not there is a node at the crossing point of two bonds from images. Consequently, image-based CNN prediction carries inherent and intrinsic limitations compared to the dynamical matrix-based prediction. By uncovering the underlying physics mechanism, we can adjust the inputs for CNN models, enabling the prediction of a wide range of Poisson's ratio values.

Finally, we compare the CNN model prediction times from image input and dynamical matrix input with the time from solving equilibrium equations, as shown in Figure 5d. Clearly, predictions based on the dynamical matrix are 25 times faster than image-based predictions, and this efficiency remains consistent across amorphous networks with varying numbers of nodes, ranging from 144 to 1024. Increasing the number of nodes amplifies the time cost of solving equations due to the rise in unknown variables. In contrast, the time cost for CNN models remains relatively constant for complex networks with varying numbers of nodes, as their hyperparameters are fixed after training. For the largest-sized amorphous network in our study, the trained AlexNet can make predictions $3.5 \times 10^3$ times faster than solving equations directly. This underscores the potential revolutionary impact of scientific discovery on the development of deep learning.

\section{Discussion and Conclusion}

We have presented a comprehensive research framework, exemplifying the entire ``AI for Science'' loop. Initially, machine learning algorithm is employed to design practical systems with various Poisson's ratio. Subsequently, the AI-generated database reveals the fundamental physical mechanism driving Poisson's ratio. This insight informs the deep learning models to utilize dynamical matrix inputs rather than image inputs, leading to a substantial increase in computational efficiency.  Note that the structures generated by our machine learning algorithm deviate significantly from previous designs, which is the crucial prerequisite for discovering novel and fundamental physical mechanisms. Traditional designs for auxetic networks relied on periodically arranged unit cells and concave structures. By contrast, our machine learning-generated structures exclusively consist of convex structures which represents a completely new design paradigm. This unique structural approach enables further exploration into the underlying mechanism of Poisson's ratio. Through normal mode analysis, we unveil that a subset of low-frequency normal modes of the dynamical matrix determines Poisson's ratio.

Given that deep learning typically relies on image datasets represented as matrices encoding RGB channel pixels, we use deep neural networks to extract mechanical properties encoded within the dynamical matrix. Additionally, traditional image-based dataset acquisition can be time-consuming and resource-intensive, making it less suitable for complex scientific inquiries. By replacing images with physical descriptors, such as the dynamical matrix, we can gain a significant efficiency advantage. This approach reduces the input training data size, as each sample now only consists of a single-channel dynamical matrix. Consequently, the training time required for a CNN model is significantly reduced. This trained CNN effectively serves as a surrogate model to avoid extensive calculations on actual systems. This breakthrough illuminates a promising pathway that bridges the gap between artificial intelligence and fundamental scientific research. Importantly, it highlights the dual role of AI, not only providing sufficient data for fundamental research but also driving the development of AI through fundamental mechanisms obtained from AI-produced data.


\section{Methods}

\subsection{Construction Methods of Amorphous Networks}

As illustrated in the left panel of Figure 1b, the intial network is a perfect triangular lattice. The disturbance vector is defined by $r$ and $\theta$, with the origin centered at each original lattice point. Here, $r$ denotes the disturbed length, which is a constant for every node. Meanwhile, each node $i$ is assigned a random $\theta_i$ within the range $[0,2\pi]$. Each node within the boundary is then distorted to a new position, given by the distortion vector expressed in Cartesian coordinates as $\delta\vec{r_i}=(r \cos{\theta_i},r \sin{\theta_i})$. After the distortion, the connections between nodes are replaced with new relaxed springs. Importantly, the topology of the constructed network, including the number and connections of springs, remains unchanged after the distortion. It is important to note that $r$ should be less than half the lattice constant to avoid bond crossings. When $r$ is above 0.5, crossing bonds start to emerge. Our simulations and experiments are always setting $r$ below 0.5 to simply the structures, except in the comparison of two types of training in Fig.5. Consider a triangular lattice consists of a total of $N_{side}\times N_{side}$ nodes, with each boundary comprising $N_{side}$ uniformly distributed nodes. As the node distortion manipulation is implemented on every node within the boundary (with the nodes on each boundary remaining fixed), a cumulative total of $N_{side}\times N_{side}-2N_{side}-2(N_{side}-2)$ nodes within each triangular network are affected by this perturbation.

Figure 1c illustrates the process of constructing a random network. In the left panel, a set of nodes (represented as balls) is randomly distributed within a rectangular region. The right panel of the figure shows the creation of bonds (relaxed springs) through Delaunay triangulation, ensuring that the mean coordination number of the amorphous network, denoted as $<z>$, is set to 6.

As shown in the left panel of Figure 1d, the network depicted is derived from a typical two-dimensional bi-dispersion packing system. This packing system consists of two types of particles with the same number, and their radius ratio is $1:1.4$. The volume fraction of these particles keep increasing until they form a jammed structure, and the contacts between neighboring particles are substituted by identical relaxed springs, as illustrated in the right panel.

\subsection{Simulated Annealing (SA) Method}

Using machine learning, we first illustrate a general and effective tuning approach for $\nu$. Although in previous studies $\nu$ can be tuned to negative in packing-derived networks \cite{goodrich2015principle,hexner2018role}, a general tuning approach for an arbitrary amorphous network is still lacking. Moreover, the previous bond-cutting procedure is based on the importance of a single bond to $\nu$, which typically leads to a locally-optimized result instead of the globally-optimized result (see Figure 1e). Thus the tuning range or efficiency is rather limited. How to approach the globally-optimized result? We tackle this problem with the machine learning algorithms developed in reinforcement learning \cite{kirkpatrick1983optimization}. Such algorithms have a feedback mechanism which modifies the optimization strategies, and allows the algorithms to accept undesirable results to `jump' out of the local minimum. Eventually the system will approach the global minimum which traps the system much better. After generating a large number of globally-optimized networks, the data set is then fed back to the actual network to get the best bond-cutting procedure for $\nu$ adjustment.

The simulated annealing (SA) algorithm is a widely used optimization method that mimics the gradual cooling process observed in metals. In each iteration, corresponding to an annealing temperature, the algorithm generates a new potential solution by modifying the current state. The new state is then accepted or rejected based on the Metropolis criteria, and this process continues until convergence. The SA algorithm is known for its ability to avoid local optima and approach the global optimal solution.

The algorithm consists of an external loop and an internal loop. The annealing process is applied in the external loop, where the system starts with an initial temperature $T_{0}$ and cools down to the next step $T_{k+1} = \alpha T_{k}$ with a cooling rate $\alpha$. $\alpha=0.95$ in our case. The annealing process terminates when the system reaches its final temperature $T_{f}$. The internal loop involves the implementation of the Metropolis principle. At each discrete temperature $T_{t}$, we iterate $L_{k}$ times to manipulate the bonds for changinge the network's structure. Here, $L_{k}$ represents the length of the Markov chain, and it represents $L_k$ manipulations at temperature $T_k$. We set $L_k=15$ in our simulation process. In each manipulation of $L_k$, a single bond could either be removed or added to the system, causing a change in the system's Poisson's ratio from $\nu(\pi_{t})$ to $\nu(\pi_{t+1})$, where $\pi_{t}$ and $\pi_{t+1}$ represent the network structures at steps $t$ and $t+1$. The probability of removing a bond is set as 0.7 at each time, which is larger than bond readding probability and ensures the overall trend is to remove bonds. In the internal loop, we introduce an acceptance probability $P$, analogous to the transition matrix used in the Markov Chain Monte Carlo (MCMC) algorithm, to facilitate the system in escaping local optima and approaching global optima.

Next, we demonstrate how the SA algorithm functions using an example of $\nu$-decreasing. After removing specific bonds, the network structure changes from $\pi_{t}$ to $\pi_{t+1}$. The acceptance probability $P$ can be represented by the following expression, which depends on the difference between $\nu(\pi_{t})$ and $\nu(\pi_{t+1})$:
\begin{equation}
    P = \begin{cases}
    1, & \text{if } \nu(\pi_{t+1}) < \nu(\pi_{t}); \\
    e^{(-\dfrac{{\nu(\pi_{t+1})-\nu(\pi_{t})}}{{T_t}})}, & \text{if } \nu(\pi_{t+1}) > \nu(\pi_{t}).
    \end{cases}
\end{equation}
Accordingly, if there is a decrease in $\nu$ after one iteration step, the operation is accepted without question ($P = 1$). Conversely, if $\nu$ increases, the algorithm considers accepting the bonds cutting strategy based on the Metropolis principle. It generates a random number $\epsilon$ within the range of [0,1] and compares it with the exponential function $P$. If $\epsilon \le P$, the strategy is accepted. Otherwise, the algorithm proceeds to the next iteration.

Visualizing the solution space as a landscape, the optimal bond cutting strategy can be considered as a ball falling into a pit. The aim of finding the globally optimal strategy is to guide the ball to find the deepest pit in the landscape. However, without a reliable algorithm, the ball may become trapped in local minima, as depicted in Figure 1e in the main text. To address this issue, the SA algorithm implements a dynamic acceptance probability determined by the temperature $T_t$ at time $t$ and the difference in Poisson's ratio between two consecutive states, $\Delta\nu = \nu(\pi_{t+1}) - \nu(\pi_t)$. By employing this dynamic acceptance probability and the Metropolis principle, the solution space is effectively stirred. This helps the ball to move out of the local minima and continue exploring the solution space until it eventually reaches the globally deepest pit. Once the ball reaches the deepest pit, it is very difficult to move out and has to stay. Similarly, for an increase in $\nu$, the acceptance probability $P$ for a bad strategy becomes $\exp((\nu(\pi_{t+1}) - \nu(\pi_t))/T_t)$ if $\nu(\pi_{t+1}) < \nu(\pi_t)$.

\subsection{Prediction of Poisson's ratio through deep learning}

The CPU utilized in our study is the Intel(R) Core(TM) i7-10700, complemented by the NVIDIA GeForce GTX 1650 GPU. We employed triangular networks with varying distortion radii to train the Convolutional Neural Network. The training dataset covered distortion radii ranging from 0.05 to 0.95 in the triangular lattice. It is noteworthy that lattice distortions exceeding 0.5 can induce a cross-bond pattern, presenting challenges for computer vision tasks to discern whether or not intersecting bonds share a common node at the crossing point. Consequently, image-based training inherently comes with a limitation. Due to the intricacies involved in image recognition, the accurate prediction requires deeper-layered neural networks. Therefore, we chose the ResNet-50 model for training with image-based datasets, as it is well-suited for handling complex datasets and can alleviate the vanishing/exploding gradient issue. Given that the dynamical matrix database encodes more precise information, such as whether two crossing bonds share a common node, we have chosen to utilize the relatively simpler model, AlexNet, for predicting Poisson's ratio using the dynamical matrix.

The dataset was partitioned into training, validation, and testing sets at a ratio of 6:2:2. Training an AlexNet model using a dynamical matrix only requires 455 samples. In contrast, training an image-based ResNet-50 model requires approximately 3640 network samples, which is almost 9 times larger than the dataset required for the dynamical matrix-based approach. For feature extraction in AlexNet, we employed five convolutional layers followed by three pooling layers for dimension reduction. This sequential process facilitates the extraction of essential information at each layer, culminating in the final pooling layer. The resulting feature map was flattened into a one-dimensional neural network, connected to two additional fully connected layers. In the context of our prediction task, the output layer consisted of a single neuron responsible for regressing the value of Poisson's ratio. The well-trained neural network serves as a surrogate model, obviating the need for matrix solving and enabling the prediction of Poisson's ratio for the given network. For post-training, we utilize the neural network to predict the Poisson's ratio for previously unseen samples constituting a separate dataset known as the testing set. The predictive performance of the optimal trained model can be quantitatively evaluated by the testing dataset with two metrics, i.e., root mean square error ($RMSE$) and determination coefficient ($R^2$) defined by: $RMSE=\sqrt{\frac{1}{N} \sum_{i=1}^N(\hat{\nu}_i-\nu_i)^2}$, $R^2=1-\frac{\sum_{i=1}^N (\hat{\nu}_i-\nu_i)^2}{\sum_{i=1}^N (\bar{\nu}_i-\nu_i)^2}$, where $\bar{\nu}_i$ denotes the average value of testing dataset, while $\nu_i$ and $\hat{\nu}_i$ denote the predicted and label Poisson's ratio values.

\noindent\textbf{Supplementary Material} See the supplementary material for the details of normal mode analysis and designed gradient materials based on the Poisson's ratio.

\noindent\textbf{Acknowledgements} L. X. acknowledges the financial support from NSFC-12074325, GRF-14307721, CRF-C6016-20G, CRF-C1018-17G, CUHK direct grant 4053582, X. S. acknowledges the financial support from NSFC, under the Grant No. 12205138, from Shenzhen Science and Technology Innovation Committee (SZSTI), under the Grant No. JCYJ20220530113206015.

\noindent\textbf{Author contributions} C. Z. performed all the machine learning computations and experiments, C. Z., X. S. and L. X. contributed to the theoretical normal mode analysis, C. F., Z. J. and B. L. helped in the experiment or the data analysis, C. Z., X. S. and L. X. prepared the manuscript, X. S. and L. X. conceived and supervised the research.

\noindent\textbf{Competing interests} C.Z., X.S., and L.X. are co-authors of a filed China patent no. 202311505603.2, which describes the methods used herein.

\noindent\textbf{Data and materials availability} All data are available in the main text or the Supplementary Information.

\noindent\textbf{Code availability} All custom computer code or algorithm used to generate results that are reported in the paper are available upon request.

\noindent\textbf{Correspondence and requests for materials} should be addressed to L. X. (xuleixu@cuhk.edu.hk) and X. S. (shenxy@sustech.edu.cn).

\clearpage
\newpage

\bibliographystyle{unsrt}
\bibliography{main_ref}

\end{document}